\documentclass[a4paper,oneside,12pt]{article}
\usepackage{graphicx}
\usepackage{dcolumn}
\usepackage{bm}
\usepackage{authblk}
\usepackage{blindtext}
\usepackage[margin=50pt]{geometry}
\usepackage{float}

\usepackage{lipsum}

\makeatletter
\renewcommand{\maketitle}{\noindent
\begin{tabular}{@{}p{\textwidth}@{}}
  \Large\textbf{\@title}\\[2ex]
  \large \@author \\[2ex]
  \@date \\[1.5em]
\end{tabular}
}
\makeatother

\begin{document}


\title{\textbf{\LARGE X-ray magnetic linear dichroism as a probe for non-collinear magnetic state in ferrimagnetic single layer exchange bias systems \vspace{1cm}}}
\author[1,2,3,*]{\bf \normalsize  Chen Luo}
\author[1]{\bf Hanjo Ryll}
\author[2,3]{\bf Christian H. Back}
\author[1,**]{\bf Florin Radu}
\affil[1]{\footnotesize Helmholtz-Zentrum-Berlin f\"{u}r Materialen und Energie, Albert-Einstein-Strasse 15, 12489 Berlin, Germany}
\affil[2]{\footnotesize Institute of Experimental and Applied Physics, University of Regensburg, 93053 Regensburg, Germany}
\affil[3]{\footnotesize Institute of Experimental Physics of Functional Spin Systems, Technical University Munich, James-Franck-Str. 1, 85748 Garching b. M\"{u}nchen, Germany}
\affil[*]{\textit{chen.luo@ur.de}}
\affil[**]{\textit{florin.radu@helmholtz-berlin.de}}


\date{\today}

\maketitle

\section*{Abstract}

Ferrimagnetic alloys are extensively studied for their unique magnetic properties leading to possible applications in perpendicular magnetic recording, due to their deterministic ultrafast switching and heat assisted magnetic recording capabilities.  On  a prototype ferrimagnetic alloy we demonstrate fascinating properties that occur close to a critical temperature where the magnetization is vanishing, just as in an antiferromagnet. From the X-ray magnetic circular dichroism measurements, an anomalous 'wing shape' hysteresis loop is observed slightly above the compensation temperature. This bears the characteristics of an intrinsic exchange bias effect, referred to as \textit{atomic exchange bias}. We further  exploit the X-ray magnetic linear dichroism (XMLD) contrast for probing  non-collinear states which allows us to discriminate between two main reversal mechanisms, namely perpendicular domain wall formation versus spin-flop transition. Ultimately, we analyze the elemental magnetic moments for the surface and the bulk parts, separately, which allows to identify in the phase diagram the temperature window  where this effect takes place.  Moreover, we suggests that this effect is a general phenomenon in ferrimagnetic thin films which may also contribue to the understanding of the mechanism behind the all optical switching effect.




\vspace{2cm}
Non-collinear magnetism is emerging as a crucially important trait of magnetic systems which are indispensable to antiferromagnetic spintronics~\cite{RevModPhys.90.015005}. Magnetic skyrmions, helical and conical states, canted spins specific to frustrated systems, domain walls in ferromagnetic and antiferromagnetic materials are all attempted to be controlled through external stimuli (electric currents, voltages, laser excitations, strain) towards functionalization for applications in modern devices. Moreover, non-collinear spin textures in ferrimagnets can give rise to anomalous Hall effect~\cite{RevModPhys.82.1539,RevModPhys.87.1213}, which enables readout in magnetic sensors. While important progress is made on the understanding of complex magnetic textures in single crystals, the miniaturization of devices requires nano-scaling of the materials, which leads to significant modifications of their bulk magnetic properties.

Among magnetic materials, rare-earth-transition-metals (RE-TM) ferrimagnetic alloys have attracted great interest because they exhibit superior flexibility in designing desired properties for ultimate functionality and as model systems for basic research in the field of spintronics.  They can be easily engineered as two ferromagnetic oppositely oriented sub-lattices in form of thin films and nanostructures with controllable perpendicular anisotropy, variable net magnetization as a function of stoichiometry and tunable spin reorientation transition temperature~\cite{radu.barriga.2018}.  They can also be assembled as heterostructures in form of spin valves and tunnel junctions~\cite{radu2012perpendicular,ijp2013122}. For example, DyCo/Ta/FeGd has been demonstrated to exhibit interlayer exchange coupling and a tunable and robust perpendicular exchange bias at room temperature, which can be set without additional field cooling cycles~\cite{radu2012perpendicular,radu.barriga.2018}. The DyCo$_{5}$ material has also been proposed to be suitable for heat-assisted magnetic recording near room temperature~\cite{PhysRevApplied.5.064007}. For the archetypical GdFeCo and other RE-TM ferrimagnets like TbFe, TbCo it has been demonstrated that their magnetization can be controlled using femtosecond laser pulses at large lateral length scales and at the nanoscale,  without applying any external magnetic field \cite{PhysRevLett.99.047601,PhysRevApplied.7.021001,alebrand.APL.2012,arora2017spatially,CARVA2017291}.  RE-TM ferrimagnetic alloys can be tuned to behave as true antiferromagnets at a  compensation temperature where the magnetic moments of the RE and TM sub-lattices are equal in size but oppositely oriented, leading to a zero net magnetization. For some RE elements with low orbital magnetic moments, comparable to the orbital magnetic moment of the TM element, the angular momentum may be quenched for a certain temperature which may lead to an acceleration of the precessional spin dynamics~\cite{PhysRevB.74.134404,Kim2017}. Ultrafast magnetization reversal across the compensation temperature of RE-TM alloys may  deterministically provide the ultimate switching speeds that can be achieved today~\cite{PhysRevLett.95.047202, PhysRevLett.99.217204, radu2011transient, ostler2012ultrafast, ADMA:ADMA201602963}.

The complex physics  near the compensation temperature is enriched by one more fascinating effect. Anomalous magnetic behavior in form of  \textit{wing shape} hysteresis loop has been reported in several RE-TM ferrimagnetic alloys that exhibit a perpendicular magnetic anisotropy such as, GdCo~\cite{1347-4065-15-S1-93}, HoCo~\cite{rata-pssa.2210620118}, TbFe~\cite{CHEN1983269}, GdFe~\cite{OKAMOTO1989259},DyCo$_{4}$~\cite{chen2015observation}, and GdFeCo~\cite{lit:amatsu-1977,lit:xu-2010,PhysRevLett.118.117203} thin films.
 When applying a magnetic field perpendicular to the sample, it is expected that the net magnetization will naturally align with the external field. However, a counter-intuitive effect is observed: the magnetization diminishes when the magnetic field overcomes a certain value, leading to a decrease or even a vanishing magnetization of the sample.

Originally, this intriguing effect was interpreted based on models assuming an alloy composition gradient across the film thickness or even across lateral directions of the sample. According to these assumptions, a compensation temperatures range will occur and, as a result the magnetic hysteresis loop will reflect the relative weight of the corresponding "below" and "above" compensation parts of the film~\cite{1347-4065-15-S1-93,lit:amatsu-1977}. These early models  have been addressed critically  in relation to similar observations in HoCo films~\cite{rata-pssa.2210620118}, questioning the original proposals based on the structural or magnetic inhomogeneities. Recently, the observation of similar anomalous loops in GdFeCo was observed to occur at faster time scales~\cite{OKAMOTO1989259}. For this case the origin for the effect was suggested to be caused by a transient temperature range which extends over the compensation temperature of the film. Even more recently, similar anomalous magnetic behavior in the same GdFeCo films was reported in equilibrium with the suggestion that its origin actually may relate to a spin-flop mechanism~\cite{PhysRevLett.118.117203}. This last attempt to resolve the debate, however, comes at odds with previous observation of this effect in a DyCo$_{4}$ film where it is suggested that the effect bears the characteristics of an exchange bias effect~\cite{chen2015observation}. As a result, although  this effect is off paramount importance for ultrafast magnetization research, its fundamental origin is still highly debated. We center our study on resolving the origin of the effect utilizing one of the most powerful modern experimental tools to magnetism, namely soft x-ray spectroscopy. Moreover, we suggests the this effect may explain the origin of all optical switching in ferrimagnetic films (see Discussion section and Supplementary).

 \section*{Results}
 
\begin{figure*}[ht]
	\centering
	\includegraphics[width=1\linewidth]{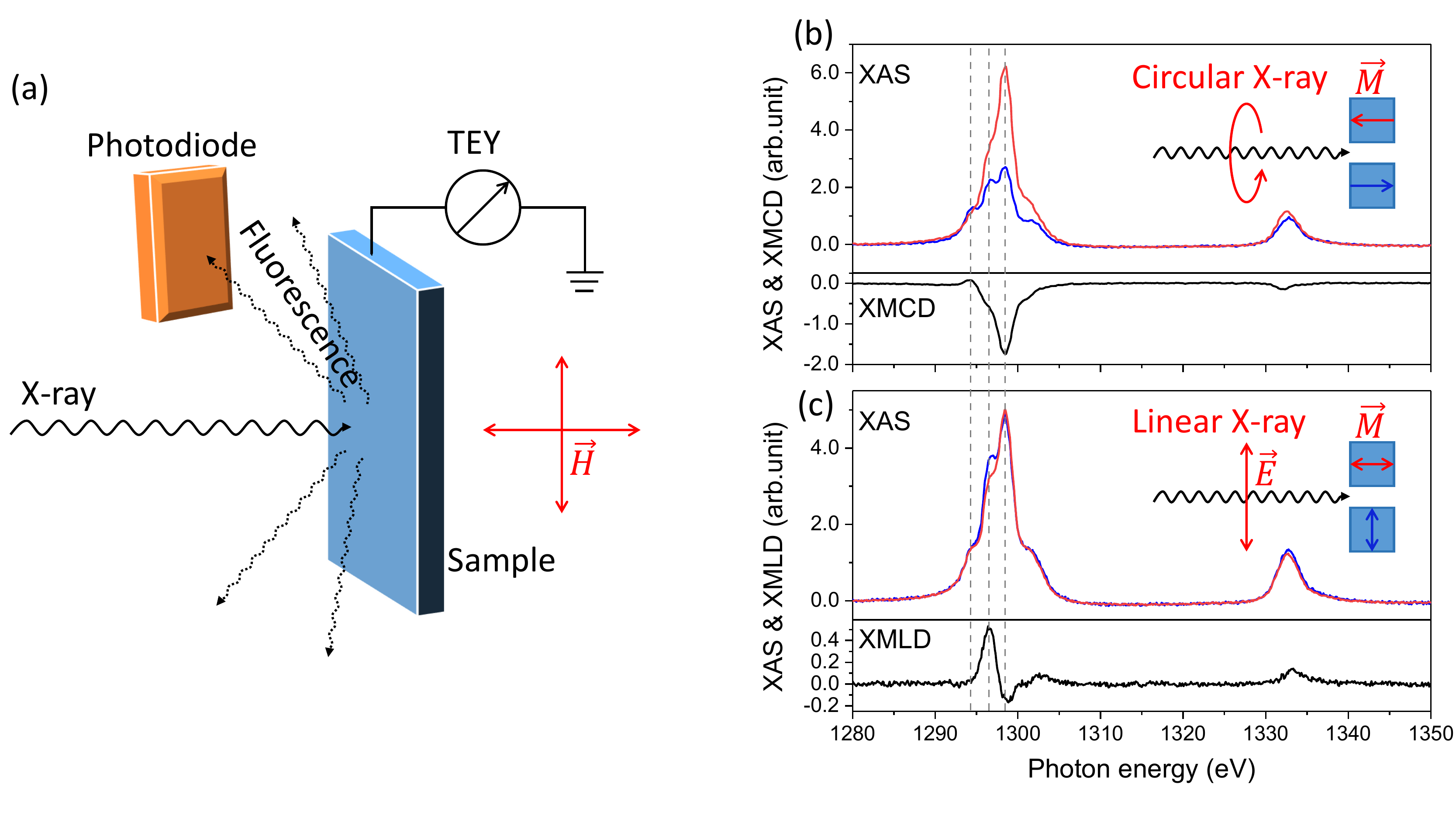}
	\caption{(a) Sketch of the XMCD and XMLD measurements, the fluorescence and TEY signals from the sample are recorded. (b) The XAS and XMCD spectra for the Dy M$_{4,5}$ edges at 300 K. (c) The XAS and XMLD spectra for the Dy M$_{4,5}$ edges at 300 K. The three peaks of Dy M$_{5}$ edge are marked by the dashed lines at $E=$ 1294.3, 1296.5 and 1298.4 eV. The data in panels (b) and (c) were both recorded in FY mode. }
	\label{fig:fig1}
\end{figure*}

We make use of x-ray circular magnetic dichroism (XMCD)~\cite{PhysRevLett.58.737} and x-ray linear magnetic dichroism (XMLD) assembled from magnetic field dependent absorption spectra (XAS) 
measured by total electron yield (TEY) and by fluorescence yield (FY) to resolve the origin of the magnetic transition that occurs close to the compensation temperature of RE-TM ferrimagnetic alloys. XMCD is sensitive to the ensemble  averaged orbital and spin contribution to the magnetic moments projected along the circular polarization direction which is set to be parallel to the x-ray beam direction. Through the sign of XMCD one can distinguish the directional sense of the magnetic moments. However, through its magnitude one cannot uniquely discriminate on their eventual non-collinear arrangement with respect to magnetic domain formation. To achieve this capability, the XMLD contrast will be involved. 

\begin{figure}[ht]
	\centering
	\includegraphics[width=1\linewidth]{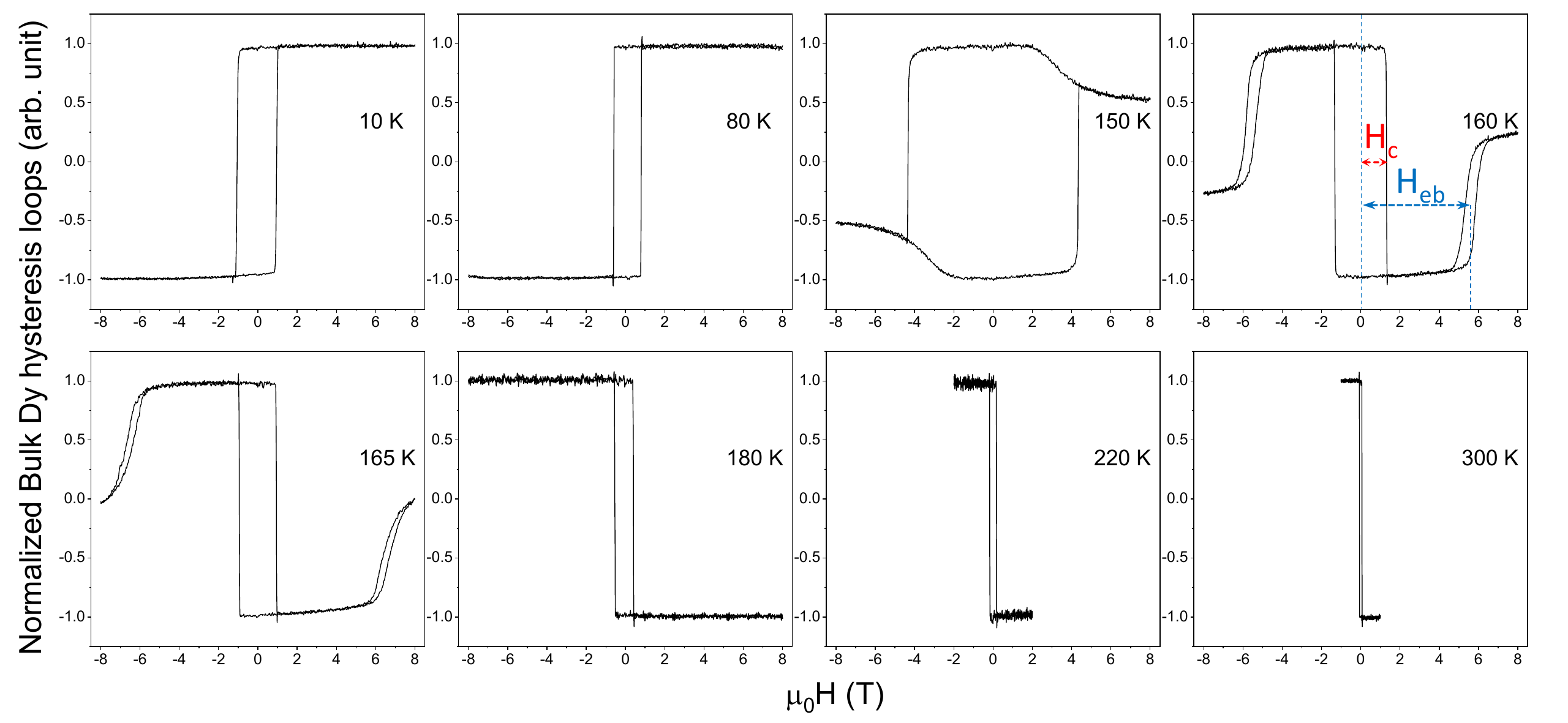}
	\caption{ Temperature dependent hysteresis loops recorded by FY signal with circular polarized X-rays set to the Dy M$_{5}$ edge ($E= 1298.4$ eV). The shift of the side hysteresis loop denoted as $H_{eb}$ and the coercive field of the central loop denoted as $H_c$ are depicted schematically in the top-right panel.
}
	\label{fig:fig2}
\end{figure}

XMLD~\cite{PhysRevLett.55.2086, PhysRevLett.70.1549, PhysRevLett.82.640, doi:10.1063/1.1456542,PhysRevLett.98.197201} is the difference in XAS cross section for the $\vec{E}$ vector of linear polarized X-rays oriented parallel and perpendicular to the magnetic moments. XMLD depends on the square of the magnetic moment $<M^2>$ and on the magneto-crystalline anisotropy,  which makes it favorable for the study of antiferromagnetic systems. In spite of the key dependences to intrinsic magnetic properties, for 3d transition metal elements (Mn, Co, Fe) the size of the XMLD effect is extremely small, hindering its further development~\cite{PhysRevLett.82.640,PhysRevB.74.094407}. Also, it requires a rather high precision of reproducibility of energy set since the size of its sign changing contrast at the L$_3$ resonant edge occurs in a narrow energy range. We exploit here, as demonstrated further below, the XMLD at the RE (Dy) edges which is expected to be much larger~\cite{PhysRevLett.55.2086} and occurs as a sizable intensity change at the M$_5$ edge for both TEY and FY detection modes.

%
First we present the demonstration of XMCD and XMLD geometries and the novel characteristics of the XMLD effect at the Dy M edges through an experiment to probe the non-collinear states in DyCo$_{5}$ ferrimagnetic thin films at room temperature. 
Figure~\ref{fig:fig1}(a) shows the experimental geometry for the surface and bulk measurements (see also Supplementary). All the measurements were performed in a perpendicular geometry, the bulk sensitive FY signal with a probing depth of $\sim$ 100 nm~\cite{ruosi2014electron} was recorded by a photodiode located 2 cm away from the sample surface. 
Fig.~\ref{fig:fig1}(b) shows the XAS and XMCD measurements for Dy at 300 K. The XMCD spectrum was obtained by taking the difference of $(\sigma^{+}-\sigma^{-})$, where $\sigma^{+}$ and $\sigma^{-}$ represent the XAS spectra measured by FY and  using the circular polarized X-rays with the magnetic field ($\mu_{0}H=1$ Tesla) parallel and anti-parallel to the beam direction. 
The Dy XMLD spectrum was obtained by taking the difference of the XAS spectra measured by keeping the linear polarized X-rays  $\vec{E}$  parallel to the synchrotron plane and perpendicular to the beam direction, and setting the direction of the magnetic moments perpendicular and parallel to $\vec{E}$ , respectively, as  shown in Fig.~\ref{fig:fig1}(c). The magnetic field of 1 Tesla was checked to saturate the magnetization of the sample whether in-plane or out-plane at room temperature. The field was applied parallel to the beam direction and perpendicular to it. As shown in the figure, there are three peaks at the Dy M$_{5}$ edge, which are marked by the dashed lines. The maximum XMCD signal appears enhanced at the third peak whereas the maximum XMLD signal is located at the middle peak. The intensity difference at the M$_{5}$ edge for the XMLD spectra is sufficiently  large to be exploited for intensity measurements as a function of the external field. The sensitivity to the angular orientation between the magnetic moments and the direction of the polarization vector is clearly demonstrated: strong XMLD contrast appears  by re-orienting the magnetic moments only, from parallel to perpendicular directions with respect to $\vec{E}$,  using vectorial  magnetic fields. To exclude further contributions to the XMLD contrast possible caused by  crystalline electric field effect,  further orthogonal field directions in plane of the sample were measured (Supplementary). 

To approach the compensation temperature, we performed temperature-dependent XMCD and hysteresis loop measurements at the Dy M$_{4,5}$ and Co L$_{2,3}$ edges.
 Part of the bulk sensitive  hysteresis loops taken  by measuring the FY intensity at the Dy M$_{5}$ edge ($E= 1298.4$ eV) are shown in Fig.~\ref{fig:fig2}. 
The hysteresis loops signal reverses when the temperature crosses a critical temperature, 
called magnetic compensation temperature T$_{comp}$. Also, the occurrence of the  perpendicular magnetic  anisotropy is clearly distinguished as the full remanent magnetization and by the sharpness of the magnetization reversal. Besides the main sharp reversal of the film, we observe an "anomalous" behavior at higher fields which develops at temperatures higher  than T$_{comp}$. This intriguing response of magnetization at higher magnetic fields is counter-intuitive in nature.  For magnetic films which exhibits a net magnetization like ferromagnets, an external field will cause full magnetization, aligning the spins as the field is increased. By contrast, the "anomalous" hysteresis loop show that the magnetization decreases as the magnetic field is increased. 

\begin{figure}[!]
	\centering
	\includegraphics[width=.5\linewidth]{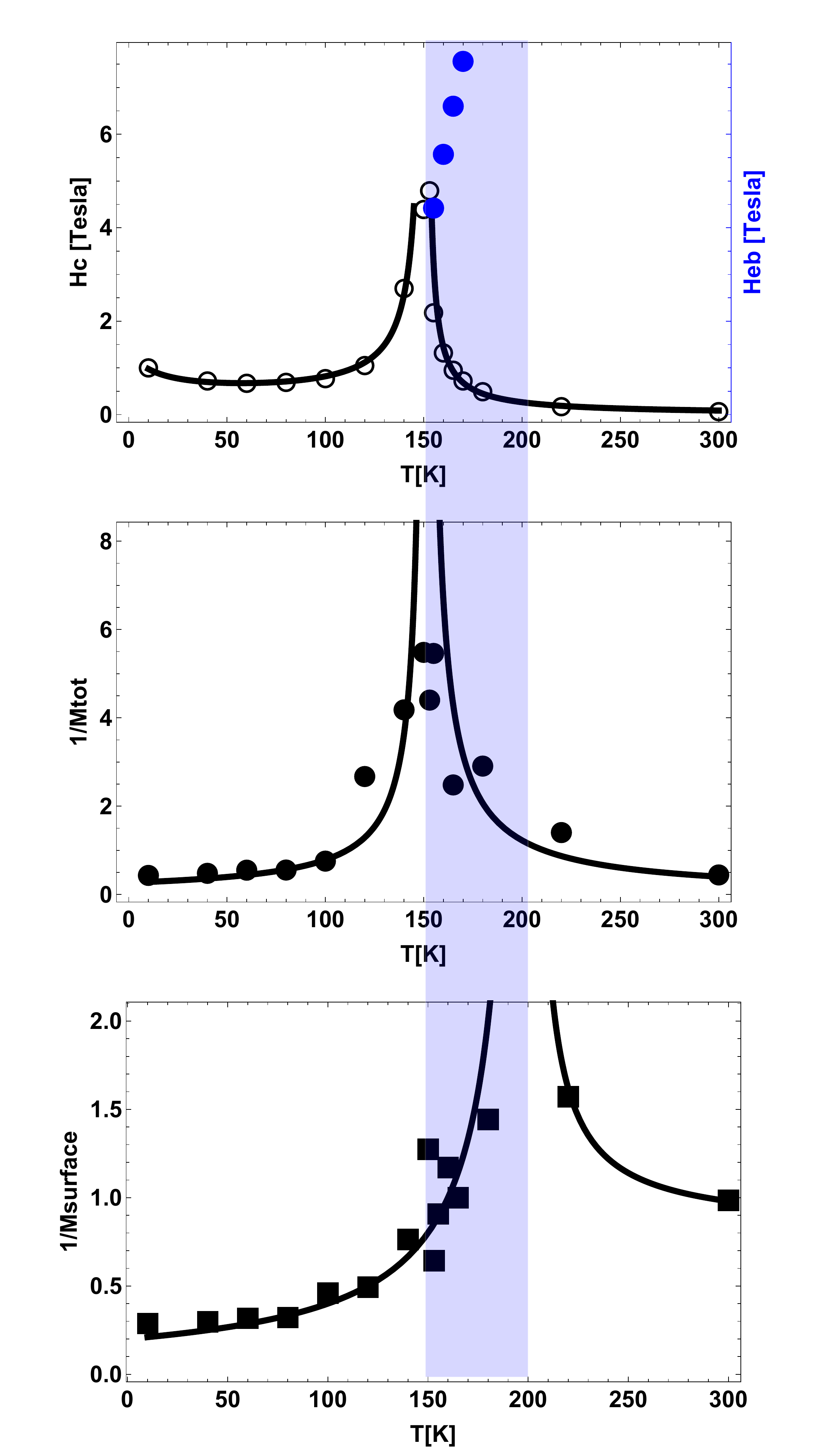}
	\caption{Phase diagram of the magnetic states as a function of temperature: (a) Temperature dependence of the coercivity field $\mu_{0}H_{c}$ and the exchange bias field $H_{eb}$. The values of $\mu_{0}H_{c}$ were extracted from the rectangular  hysteresis loops at the crossing point with respect to the magnetization axis. The shift of the side wings, denoted as $H_{eb}$, were extracted at the half height of the side loops alone. The maximum value of $\mu_{0}H_{c}$ is about 4.8 T slightly below T$_{comp}$, whereas the highest measured shift side  hysteresis loop is about 7 Tesla; (b) The inverse of the total net magnetic moment extracted by analyzing the XMCD spectra characteristic for the bulk part of the film (FY data). This show a divergent behavior at the closely similar compensation temperature as in panel (a). (c)  The inverse of the total net magnetic moment extracted by analyzing the XMCD spectra characteristic for the surface part of the film (TEY data). They also exhibit a divergent behavior near 200 K, showing the probed surface has a higher compensation temperature. In between these two compensation temperatures, the side hysteresis loops occur. }
	\label{fig:fig2b}
\end{figure}

 In Fig.~\ref{fig:fig2b}(a) we plot the coercive field of the hysteresis loops and the shift of the side hysteresis loops as a function of temperature. The divergence of the coercive field, which occurs at the vanishing net magnetization of the film, reveals with good accuracy the absolute value of the intrinsic T$_{comp}$ which is about 154 K, in close agreement with previous experiments and simulation results ~\cite{TSUSHIMA1983197,PhysRevB.96.024412}. Also, we plot the field of the center of the wing hysteresis loops as a function of temperature, denoted as exchange bias field H$_{eb}$. We observe that the shift of the side loop increases as the temperature increases up to the highest measured value of about 7~T. Due to the finite available external fields (up to 9~T) we could not further follow the shift of the side loop beyond the 7~T. This limitation is visible  at 165~K, where the side loop begin to behave as a minor hysteresis loops. Instead, we can determine the limiting temperature where the side loop will vanish. To this end we show in Fig.~\ref{fig:fig2b}(b) and (c)  the inverse of the total net magnetic moments characteristic of bulk and surface, respectively. They are extracted from XMCD spectra measured by TEY and by FY (Supplementary). We observe that both the inverse of the bulk and the surface net magnetic moments exhibit a divergent behaviour, at 154~K for the bulk part of magnetization and at 200~K for the surface part. 
As such, the system exhibits a different compensation temperature for the probed surface, which is about 50~K higher as compared to the bulk magnetic compensation~\cite{chen2015observation}. In-between these two compensation temperatures the system is in a frustrated state leading to the peculiar wing shape hysteresis loops. The occurrence of these two compensation temperatures correlates very well with the temperature range where the anomalous magnetic behaviour occurs at high fields. Outside this region, we observe no side hysteresis loops because the surface and the bulk spins are in a stable configuration. 


\begin{figure}[!]
	\centering
	\includegraphics[width=1\linewidth]{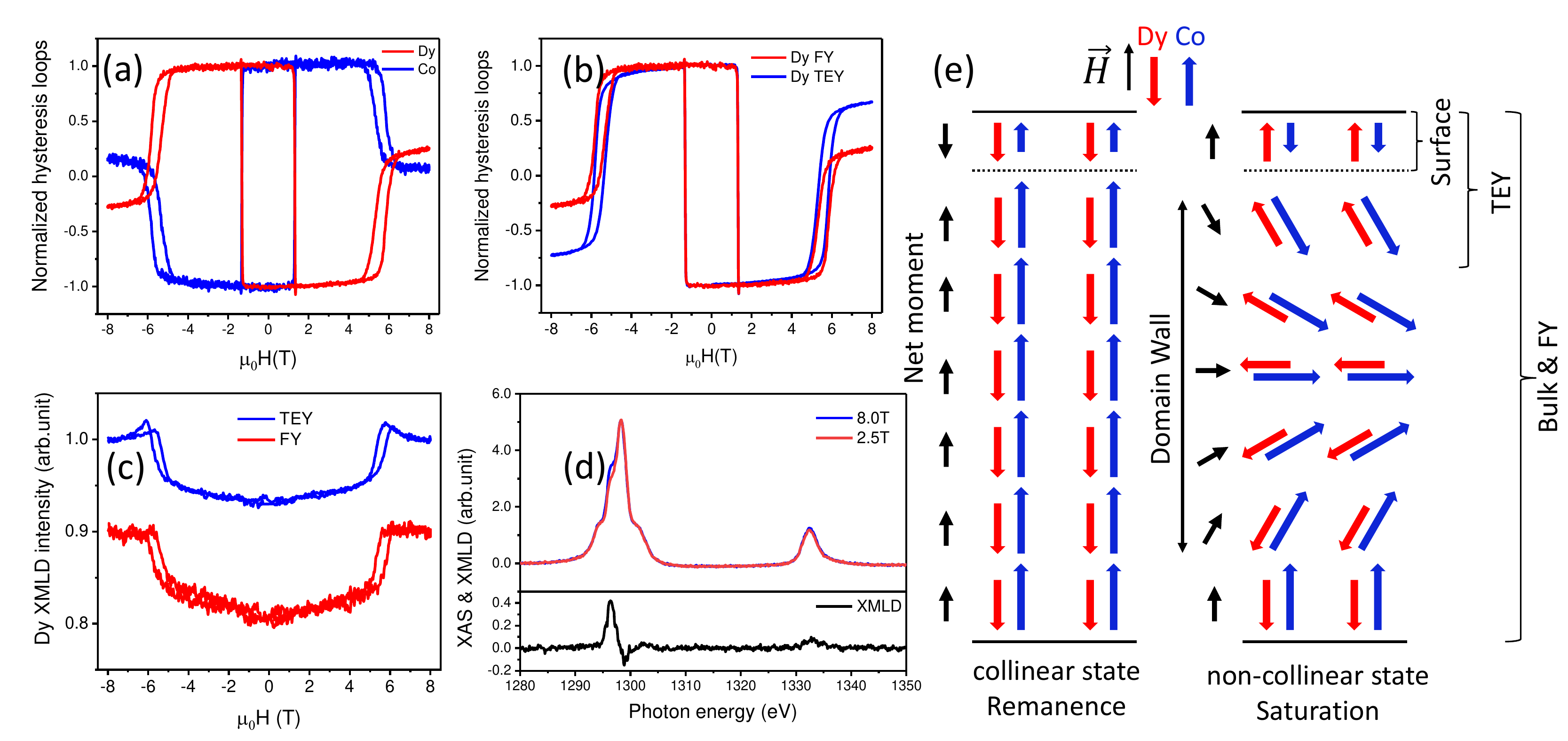}
	\caption{XMCD and XMLD measurements at 160K: (a) Hysteresis loops measured by recording the FY signal with \textbf{\textit{circular}} polarized X-rays at the third peak of the Dy M$_{5}$ edges ($E= 1298.4$ eV) and the Co L$_{3}$ edge ($E= 777.0$ eV). (b) Comparison of surface (TEY) and bulk (FY) hysteresis loops for Dy. (c) XMLD hysteresis loops measured by recording the FY and TEY signal with \textbf{\textit{linear}} polarized X-rays at the second peak of the  Dy M$_{5}$ edges ($E= 1296.5$ eV).  (d) XAS and XMLD spectra taken at 8 T and 2.5 T for Dy with linear polarized X-rays. (e) Sketch of the magnetic spin structure for the remanence and saturation state.}
	\label{fig:fig3}
\end{figure}

To characterize the reversal of the anomalous loops we make use of the XMCD and XMLD effects,  focusing on the relevant temperature of 160 K. Fig.~\ref{fig:fig3}(a) shows the hysteresis loops at both Dy M$_{5}$ edge and Co L$_{3}$ edge. They exhibit a similar behaviour with opposite signs. This demonstrates that the magnetic moments of the Dy and Co sublattices are basically anti-parallelly coupled to each other, even when they enter the anomalous spin state at higher fields. At this stage, we can resolute that a spin-flop transition does not occur. Such a spin-flop would appear in the Fig.~\ref{fig:fig3}(a) as a significant difference between the shape of the magnetization curves of Dy and Co, which is not observed. 
However, a weak non-collinear state between Co and Dy net magnetic moments can be distinguished by the  difference of the relative  magnetizations present at $\pm$8 T. This observation  is further  supported by the demonstration of the same effect in a FeGd film (Supplementary). 
We can consider that the Co and Dy sublattices are essentially anti-parallelly oriented for all external magnetic fields. This, however, does not exclude a non-collinear behavior  for the elemental sublattices:  comparing the surface and bulk hysteresis loops, see Fig.~\ref{fig:fig3}(b),  clearly reveals that the surface signal is almost completely reversed at high fields while the bulk signal is only half reversed, which strongly indicates that the mechanism is due to the reversing of the surface magnetic moments.
 
By taking advantage of the strong linear dichroism of the Dy element at the M$_{5}$ absorption edge, XMLD measurements were applied to understand the anomalous magnetic behavior. Here, we use linear polarized X-rays to perform  hysteresis loop measurements, as shown in Fig.~\ref{fig:fig3}(c). The XMLD hysteresis loops were recorded at the middle peak of the Dy M$_{5}$ edges at $E= 1296.5$ eV, where we observed a maximum XMLD signal in Fig.~\ref{fig:fig3}(d) and Fig.~\ref{fig:fig1}(c).  From Fig.~\ref{fig:fig3}(c) one can see that there are two hysteresis-loop-like structures for both surface and bulk. 
The two loops appear at the same field of the XMCD 'wing shape' hysteresis loops but end up at the same level of intensity at $\pm$8 T. 
These results reveal that parts of the magnetic moments rotate from the out-of-plane to the in-plane direction at high fields, which directly indicates the existence of 
non-collinear spin structure
between the surface and the bulk. 
One needs to point out that there are also significant differences between the bulk and surface signals. The surface XMLD hysteresis loop reaches a maximum value around $\pm$6 T, indicating that the surface magnetic moments rotate to the in-plane direction around this field, then decreases and approaches flattening near $\pm$8 T. This further  indicates  that the surface magnetic moments have a higher rotation angle at high fields. This is in full agreement with a simple domain wall structure initiated at the surface of the film.

Figure~\ref{fig:fig3}(d) shows the XAS spectra of Dy and their difference taken at 8 T and 2.5 T with linear polarized X-rays. By comparing the image with the standard XAS and XMLD measurements at 300 K, one can see that the XAS at 8 T is similar to the $\sigma_{\parallel}$  XAS while the XAS at 2.5 T is close to the $\sigma_{\perp}$ XAS in Fig.~\ref{fig:fig1}(c). By defining the relative XMLD amplitude as ($\sigma_{\parallel}$-$\sigma_{\perp}$)/($\sigma_{\parallel}$+$\sigma_{\perp}$), we get a value of $\sim$4.4\% here and $\sim$5.6\% at 300 K. Note that the XMLD amplitude is believed to be proportional to the square of the total magnetic moments $M^2$~\cite{PhysRevLett.55.2086}. The magnetic moments at 160 K are about $m_{bulk}^{160K}=m_{s}+m_{l}=6.8\mu_{B}$/atom and $m_{surface}^{160K}=5.7\mu_{B}$/atom, which are about 1.45 times of the magnetic moments at 300 K $m_{bulk}^{300K}=4.8\mu_{B}$/atom and $m_{surface}^{300K}=3.8\mu_{B}$/atom. Thus the relative XMLD amplitude of $5.6\%\times1.45^2=11.6\%$ can be expected at 160 K for the situation that all the magnetic moments align in-plane versus out-of-plane. Here the experimental value of 4.4\% means that the in-plane contribution at 8 T is about $\sqrt{4.4\%/11.6\%}\approx62\%$ of the total magnetic moment, which indicates a very thick domain wall probably throughout the whole film.

  \section*{Discussion}
Based on the experimental facts, one can draw a sketch of the spin structure for the anomalous magnetic behaviour, as shown in Fig.~\ref{fig:fig3}(e). The surface magnetic moments are always smaller than the bulk magnetic moments for both Dy and Co. At the magnetic remanence, the spins of Dy and Co are in a collinear state and anti-parallelly coupled to each other. Due to the strong exchange coupling between the surface and bulk, and due to the fact that the surface magnetism is dominated by Dy while the bulk magnetism is dominated by Co, the net moments of the surface would prefer to align anti-parallel towards the bulk moments. At high fields, this frustrated state becomes unstable,  which forces the spin structure of the whole system to turn into a non-collinear state with an out-of-plane partial domain wall. Within this wall, the exchange energy is stored and released, which resembles closely exchange bias interactions with the difference that the atomic exchange is not affected by additional interfaces in an otherwise  generic "two magnetic layers" system. By analyzing the net magnetic moments for the bulk and surface, separately, one can localize the temperature range where the occurrence of the side loop takes place. Within this temperature range the shift of the side loop increases as a function of temperature. This can be understood within the general theories for exchange  bias~\cite{Radu2008}, which postulates the shift of the hysteresis loop is inverse proportional to the $M \times t$, where M is the magnetization of the active magnetic layer and t is its thickness. In our case the active layer is the surface which has a compensation temperature of 200 K, therefore, when approaching this temperature the shift of the hysteresis  should increase, as observed experimentally in Fig.~\ref{fig:fig2b}.

 To suggest that this effect is a general phenomenon which occurs in thin ferrimagnetic films close to the compensation temperature, we provide supplementary data  (Supplementary)  on yet another system, namely FeGd ferrimagnetic film. There, the same effects are observed. Nevertheless, since FeGd has a lower magnetic anisotropy and stiffness (due to the nearly vanishing moment of Gd), the temperature range where the shift of the hysteresis loop occurs is larger.

 At a more general level, we speculate that our observations may have an impact towards deeper understanding of key aspects of ultrafast magnetic switching in ferrimagnetic films~\cite{PhysRevLett.99.047601,radu2011transient}. The mere occurrence of two compensation temperatures comprising two magnetic states in a frustrated arrangement should motivate further experiments which includes this peculiar phase diagram, also considering thickness~\cite{Arora2017} as a control parameter. As an alternative mechanism for the all optical switching, measured from below compensation in an external field, we can assume an intersection between the  intrinsic dynamical paths and the static ones. For instance, if the $H_{eb}$ will scale down in the non-equilibrium state during the pump probe delay, one can transiently cross over the formation of the domain wall state leading to the so called transient-ferromagnetic state observed for an FeCoGd film in~\cite{radu2011transient}. In fact, strong evidence for this scenario is seen in the Figure S5(a) (Supplementary). There, we observe that  the regions between the
-6 T and -2 and between 2 T and 6 T are very similar to the so called transient ferromagnetic-like
state observed in all optical switching experiments on FeCoGd samples. The projection of the Gd magnetization on the  field direction is opposite (after a crossing field equal to 3.5~T) with respect to the magnetization at magnetic remanence, whereas the projection of  Fe sublattice magnetization is vanishing. As such, the total net magnetic moment of the whole film is oppositely oriented at high fields (larger than 3.5~T) with respect to the net magnetic moment in lower applied fields (smaller then 3.5~T). Thus, assuming that during the pump probe delay, the time-dependent non-equilibrium states causes the system to cross the critical field for the formation of this "spring" spin configuration, one can understand the origin of the  all optical switching in ferrimagnetic thin films by analogy to the effect we reported here. Note, that this mechanism excludes the occurrence of the same all optical switching effect in ferromagnetic materials, because  this magnetic spring does  not occur in these materials. Instead, the thermally assisted all-optical switching may (disjunctively from the effect we report) take place in ferromagnetic as well as in ferrimagnetic materials as considered in~\cite{Mangin2014, hassdenteufel}.

In conclusion, DyCo$_{5}$ ferrimagnetic thin films were investigated with XMLD and XMCD techniques. An anomalous 'wing shape' hysteresis loop, referred to as atomic exchange bias effect with a large exchange bias field of $\mu_{0}H_{EB}$ up to a maximum measurable value of 7 Tesla, was observed slightly above the compensation temperature T$_{comp}\approx154$ K.
The origin of this effect which is demonstrated to be mediated by the formation of an out-of-plane partial domain wall during the hysteresis measurements, is directly confirmed via XMLD measurements.  Such a huge perpendicular exchange bias effect in a single film may be a good candidate for future perpendicular magnetic recording applications. The technique of using the XMLD contrast at the rare earth M$_{4,5}$ edges to probe the non-collinear states could be very useful for characterization of non-collinear magnetism which is intimately related to the spintronics research field.

\section*{Methods}

\subsection*{Sample preparation}The 20 nm thick DyCo$_{5}$ thin films were grown on sapphire substrates by magnetron sputtering (MAGSSY chamber at HZB) in an ultra-clean Argon atmosphere of $1.5\times10^{-3}$ mbar with a base pressure of $<2\times10^{-8}$ mbar at room temperature. The stoichiometry of the ferrimagnetic alloy was controlled by varying the deposition rate of Co and Dy targets in a co-evaporation scheme. A 3 nm Ta capping layer was grown on top of the samples to prevent surface oxidation. We have characterized the lateral  homogeneity using energy dispersive scanning  X-Ray spectroscopy technique (EDS), but were unable to observe any phase separation at the sensitivity level of about few hundred nanometers provided by this method.

\subsection*{X-ray absorption spectroscopy}The XAS measurements were performed at the VEKMAG end-station~\cite{RaduNoll2017} installed at the PM2 beamline of the synchrotron facility BESSY II. This end station offers unique capabilities for this type of research, since it provides a vector magnetic field option with a maximum magnetic field up to 9 Tesla in the beam direction, 2 Tesla in the horizontal plane and 1 Tesla in all directions for a temperature range of 2 K - 500 K. The XAS spectra  were recorded by means of total electron yield (TEY) and by fluorescence yield (FY), and for each constituent element, separately. 

The TEY is measured by recording the drain current as a function of the x-ray photon energy normalized by a Pt grid x-ray monitor mounted  in a magnetically shielded environment as the last optical element before the sample. The TEY is known to be surface sensitive, providing information over the escape length of the electrons which exhibits a mean free path of about 3 nm. As such, the surface magnetic properties are provided in a selective manner by this recording channel. 

The FY is measured by a magnetically insensitive x-ray detector, placed at 2 cm away from the sample surface. FY is a photon-in photon-out spectroscopic technique which provides information integrated over the penetration depth of the x-ray, which can be of order of tens of nm. The depth sensitivity depends on the photon energy, absorption cross-section, and on the stoichiometry of the film. As such, the FY provides magnetic information for the whole film thickness, which we denote as "bulk" sensitive. The spectra are recorded as function of the x-ray energy and normalized by the same magnetically shielded x-ray monitor.

\bibliographystyle{unsrt}
\bibliography{DyCo5}

\section*{Acknowledgments}
The x-ray absorption measurements were carried out at the
VEKMAG end-station installed at the PM2-VEKMAG beamlines, of BESSY II, Helmholtz-Zentrum Berlin (HZB). We thank the HZB for the allocation of
synchrotron radiation beamtime. The authors acknowledge the financial support for the VEKMAG project and for the PM2-VEKMAG beamline by the German Federal Ministry for Education and Research (BMBF 05K10PC2, 05K10WR1, 05K10KE1) and by HZB. Steffen Rudorff is acknowledged for technical support.
\end{document}


\title{\textbf{Suplementary information for:\newline \newline  \LARGE X-ray magnetic linear dichroism as a probe for non-collinear magnetic state in ferrimagnetic single layer exchange bias systems \vspace{1cm}}}
\author[1,2,3,*]{\bf \normalsize  Chen Luo}
\author[1]{\bf Hanjo Ryll}
\author[2,3]{\bf Christian H. Back}
\author[1,**]{\bf Florin Radu}
\affil[1]{\footnotesize Helmholtz-Zentrum-Berlin f\"{u}r Materialen und Energie, Albert-Einstein-Strasse 15, 12489 Berlin, Germany}
\affil[2]{\footnotesize Institute of Experimental and Applied Physics, University of Regensburg, 93053 Regensburg, Germany}
\affil[3]{\footnotesize Institute of Experimental Physics of Functional Spin Systems, Technical University Munich, James-Franck-Str. 1, 85748 Garching b. M\"{u}nchen, Germany}
\affil[*]{\textit{chen.luo@ur.de}}
\affil[**]{\textit{florin.radu@helmholtz-berlin.de}}


\date{\today}

\maketitle


\section*{\Large Supplementary Notes}
\section{\large XMLD: experimental geometries utilizing vectorial magnetic fields}

\begin{figure}
	\centering
	\includegraphics[width=1\linewidth]{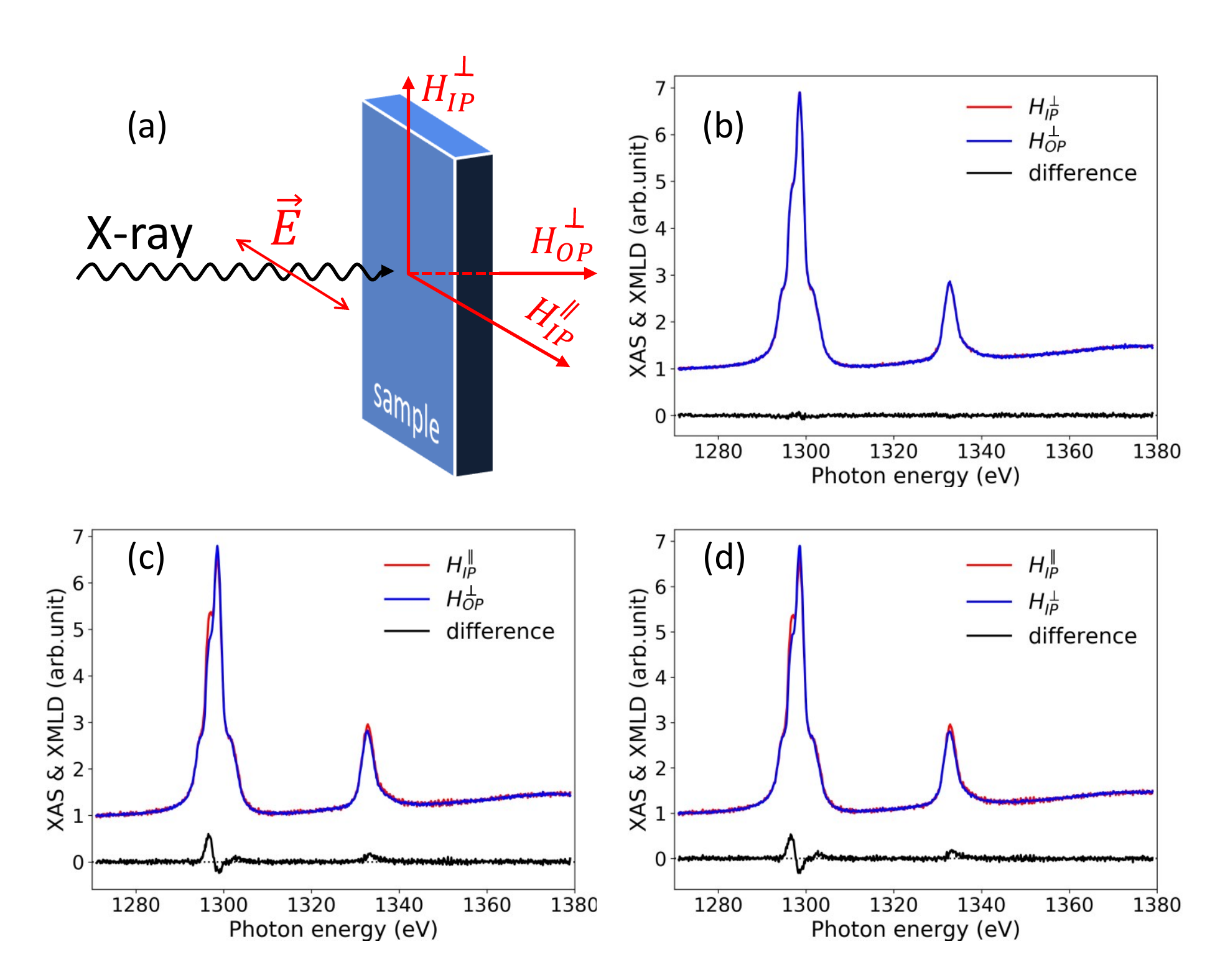}
	\caption{(a) Sketch of the XMLD measurements at room temperature. Here $H_{IP}^\parallel$ represents the field direction which lies in-plane and parallel to the $\vec{E}$ vector of the linear polarized X-rays, $H_{IP}^\perp$ represents the field direction which lies in-plane and perpendicular to $\vec{E}$, and $H_{OP}^\perp$ is the out-of-plane direction (perpendicular to the $\vec{E}$). (b,c,d) show the XAS spectra measured at $H_{IP}^\parallel$, $H_{IP}^\perp$ and $H_{OP}^\perp$, as well as their differences. The XAS spectra were measured by recording the FY as a function of x-ray energy across the $M_5$ and $M_4$ edges of Dy.}
	\label{fig:figs1}
\end{figure}

We describe below the experimental geometries for the linear dichroism  measurements using vector magnetic fields on the Ta/DyCo$_5$/A$l_2$O$_3$ sample of the main body of the paper. This demonstrates in a self-consistent manner the sensitivity of the XMLD contrast to the angle of magnetization direction with respect to the direction of the linear polarization. There are two common ways to measure a XMLD spectra. One is to keep the magnetization along the easy direction and to rotate the polarization direction, the other one is to keep the linear polarization direction fixed and change the magnetization direction~\cite{PhysRevLett.82.640}. Here, by taking  advantage of the 3D vector magnet of the VEKMAG end-station, the second method is being used for our measurements. We set the linear polarization $\vec{E}$ oriented perpendicular to the beam direction and parallel to the storage ring  plane. The sample is  oriented perpendicular to the beam direction, as for transmission geometry. The magnetization of the sample is set oriented by  the external field in three orthogonal directions: $H_{IP}^\parallel$, $H_{IP}^\perp$, and  $H_{OP}^\perp$ , as shown in Fig.~\ref{fig:figs1}(a). The $OP$ and $IP$ indexes refer to the direction of the magnetization with respect to the sample, namely out-of-plane and in-plane, respectively.  The $\parallel$ and $\perp$ indexes refer to the orientation of the magnetization with respect to the linear polarization direction of the X-rays, namely parallel or perpendicular to $\vec{E}$, respectively. 

The XAS spectra and their difference are shown in Fig.~\ref{fig:figs1}(b,c,d). The difference between the spectra measured for the orthogonal directions $H_{IP}^\parallel$ and $H_{OP}^\perp$, as well as for $H_{IP}^\parallel$ and $H_{IP}^\perp$, do show a XMLD contrast (black line in Fig.~\ref{fig:figs1}(c,d)). By contrast, when we compare the spectra recorded for  $H_{IP}^\perp$ and $H_{OP}^\perp$, we observe a vanishing XMLD contrast (black line in Fig.~\ref{fig:figs1}(b)). This demonstrates that the XMLD contrast is sensitive only to the orientation between the direction of magnetization with respect to the direction of the linear polarization. Given that the amplitude of the XMLD is large, reaching a $\sim5.6\%$ at room temperature, the intensity difference at the middle peak of the $M_5$ edge, allows for detecting non-collinearity between the magnetization and the direction of the external field during hysteresis measurements, as described in the main body of the paper.

\newpage

\section{\large XMCD:  analysis of the magnetic moments}
\begin{figure}
	\centering
	\includegraphics[width=1\linewidth]{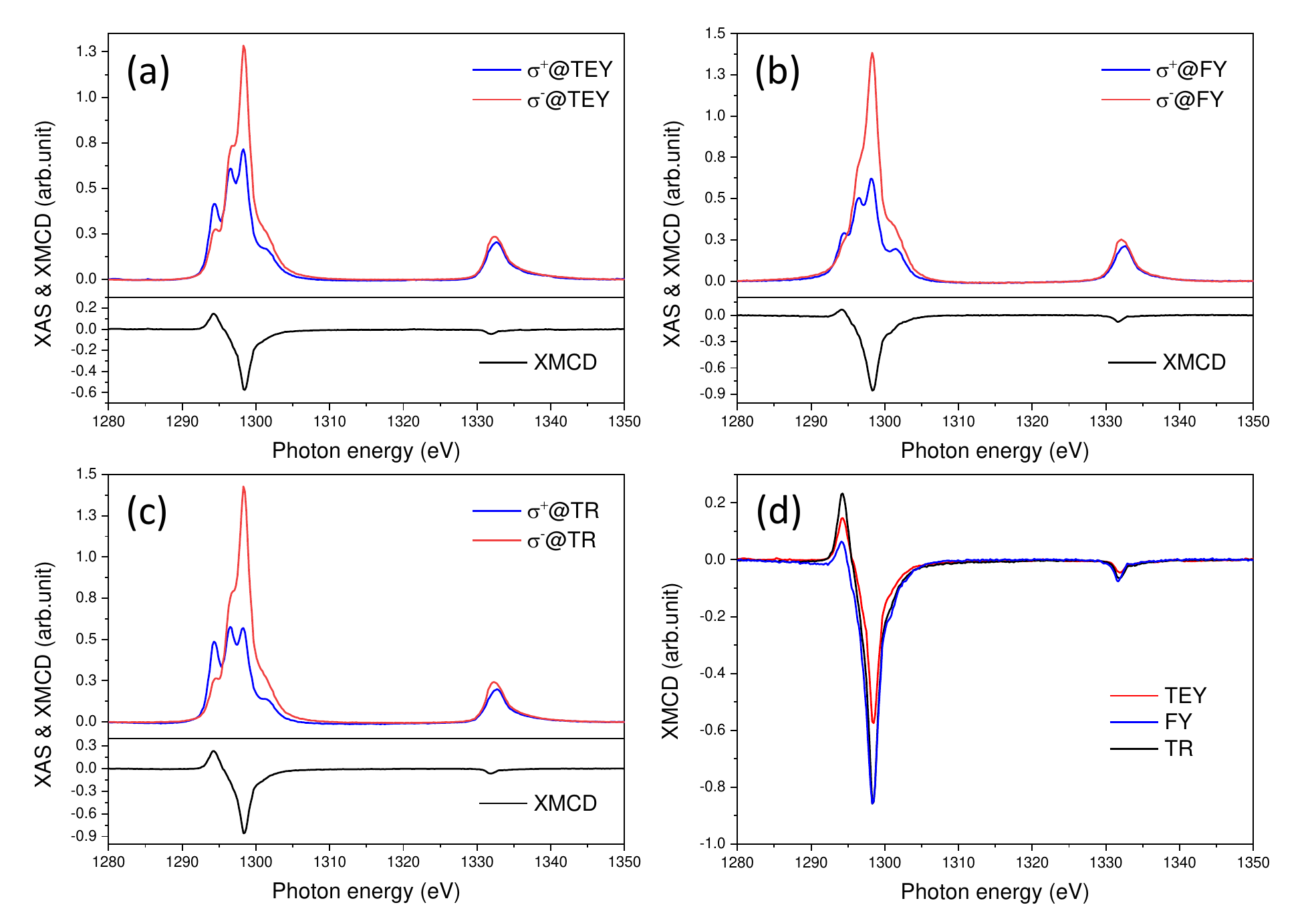}
	\caption{The Dy XAS and XMCD spectra measured by recording the TEY (a), FY (b) and transmission signals (c). (d) Comparison between the TEY, FY and transmission XMCD spectra.}
	\label{fig:figs2}
\end{figure}

Due to the existence of the self-absorption and radiative decay effects, the FY spectra may exhibit a nonlinear dependence with respect to the absorption cross section~\cite{,PhysRevB.47.14103,PhysRevB.56.2267}. To compare the difference between the XMCD measured in FY, TEY and transmission mode, we prepared a calibration sample  Ta(2.5~nm)/DyCo$_5$(20~nm)/Ta(5~nm)  grown on a 150 nm thick Si$_3$N$_4$ membrane substrate. This allows us to measure the XAS and XMCD using TEY, fluorescence and transmission signals at the same time, as shown in Figure~\ref{fig:figs2} (a,b,c). For a direct comparison, we have re-plotted all three  XMCD spectra in Figure~\ref{fig:figs2} (d).  Comparing the XMCD spectra measured in transmission with the XMCD spectra measured in TEY mode, we observe that they  have an identical line shape, with a clear difference in amplitude. This directly demonstrates that the magnetic moments of the surface are smaller as compared to the magnetic moments of the bulk part. However, the magnetic moments of the bulk measured by transmission and FY should have the same amplitude, which should be reflected in a similar amplitude and line-shape for the XMCD spectra. This is, however, not the case as directly demonstrated in Figure~\ref{fig:figs2} (d): the amplitude of the XMCD spectra is similar for the transmission and FY modes, but the XMCD line-shape measured by FY exhibits an enhanced spectral feature at ~1294.3 eV. This difference prevents an accurate determination of  the spin and orbital moments through FY measurements. Nevertheless, we have applied the sum rules  to the FY and transmission XMCD spectra and obtained the following magnetic moments at room temperature:  $m_{total}^{FY}=4.8\mu_{B}$/atom and $m_{total}^{TR}=3.9\mu_{B}$/atom. One can see that the magnetic moment obtained from the fluorescence spectra is about 23\% larger as compared to the value from the transmission spectra. 

\begin{figure}
	\centering
	\includegraphics[width=.5\linewidth]{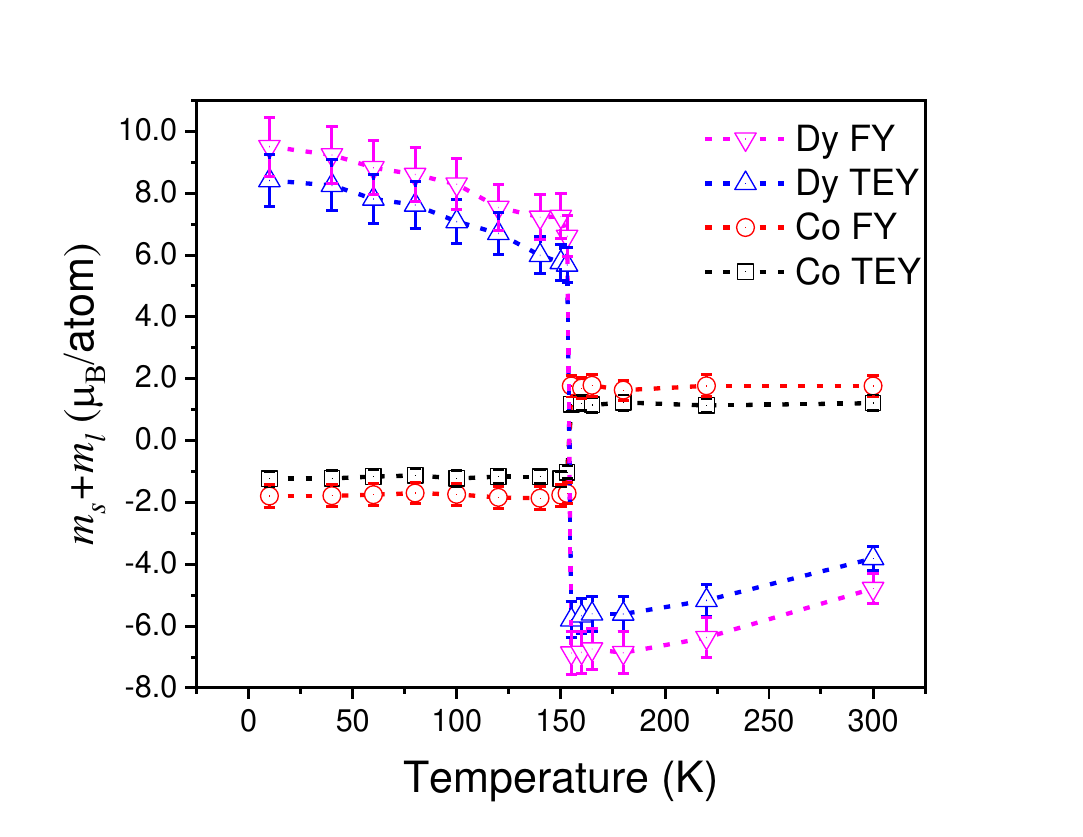}
	\caption{Temperature dependence of the total magnetic moments $m_s+m_l$. The spin $m_s$ and orbital $m_l$ magnetic moments were obtained by applying the XMCD sum rules~\cite{PhysRevLett.70.694,PhysRevLett.68.1943,PhysRevLett.75.152} for the XAS and XMCD spectra measured at remanent magnetization. The sign of the magnetic moments reverses after crossing T$_{comp}$.}
	\label{fig:figs2TD}
\end{figure}

 We then further use this scaling factor to correct for the magnetic moments extracted by applying the sum rules to the temperature dependence of the Dy XMCD spectra measured by FY in the main paper.  For XMCD measured by FY for Co (not shown), this factor turned out to close to 1. We observe that both Co and Dy exhibit lower moments on the surface for the whole temperature range, with different percentages: the Co surface has about 67\% of the bulk net moment value, whereas the Dy surface has about 84\% of its bulk value. The fact that the Co moment is reduced stronger as compare to the Dy moment  may indicate a non-collinear arrangement between the Co and the Dy spins at the top surface. This can be a result of a reduced surface coordination. The results are plotted in Fig. \ref{fig:figs2TD} to serve as the basis for the observed difference between the compensation temperatures of bulk and surface parts of the film.

\newpage
\section{\large Support study of the atomic exchange bias in FeGd film}

For the purpose of supporting our results and method, and to demonstrate a general character of our observations, we show one more model system, namely a FeGd ferrimagnetic thin film. The sample has the following structure Ta(2.5~nm)/Fe$_{77}$Gd$_{23}$ (20~nm)/Ta(5~nm)/Si$_3$N$_4$. The Gd layer exhibits a nearly vanishing orbital magnetic moment, therefore the magnetic anisotropy and the stiffness of the film are much lower as compared to the Dy based alloys. As such it can be used  as soft magnetic element in ferrimagnetic spin valves and it is the preffered system for the all optical ultrafast magnetic switching effect. 

\begin{figure}
	\centering
\includegraphics[width=.8\linewidth]{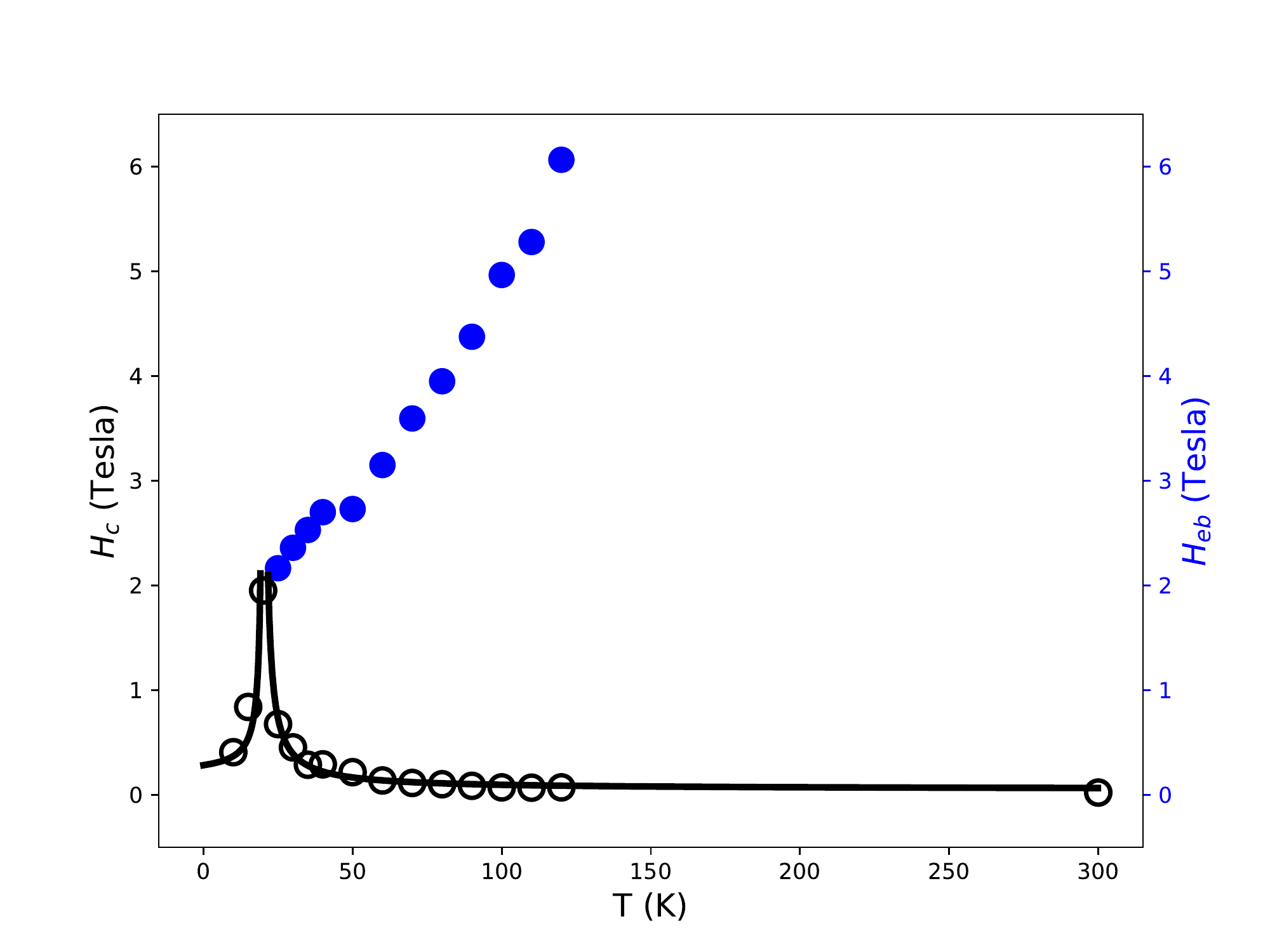}
\caption{Temperature dependence of the coercive field $H_{c}$ (open black circles) and the exchange bias field $H_{eb}$ (filled blue circles) for an FeGd film. The values of $H_{c}$ were extracted from the rectangular  hysteresis loops at the crossing point with respect to the magnetization axis. The shift of the side wings, denoted as $H_{eb}$, were extracted at the half height of the side loops alone. The maximum value of $H_{c}$ is about 2 T slightly aside T$_{comp}$, whereas the highest measured shift of the side hysteresis loop is about 6.2 Tesla. The lines are guides to the eyes.}
	\label{fig:FeGdPD}
\end{figure}

In Figure~\ref{fig:FeGdPD} we show the dependence of the coercive field and the shift of the side loop as a function of temperature. The coercive field exhibits a typical divergent behaviour at the compensation temperature which for this sample is about 25~K.  Above the compensation temperature, the system develops an additional hysteresis loop denoted as H$_{eb}$ which increases steadily up to a maximum measured value of about 6.2~T. Comparing this phase diagram with the one of DyCo$_5$ film we notice a similar character, but with some markedly differences: the temperature range where the atomic exchange bias occur is larger for FeGd (more 100 K) as compared to DyCo$_5$. Also, the temperature dependence of the H$_{eb}$ shows deviations from a linear behaviour which is actually expected within the theories of exchange bias effect. Note that for the FeGd samples we did not measure the magnetic moments, as such we cannot provide the exact compensation temperature of the surface part.

\begin{figure}
	\centering
	\includegraphics[width=1\linewidth]{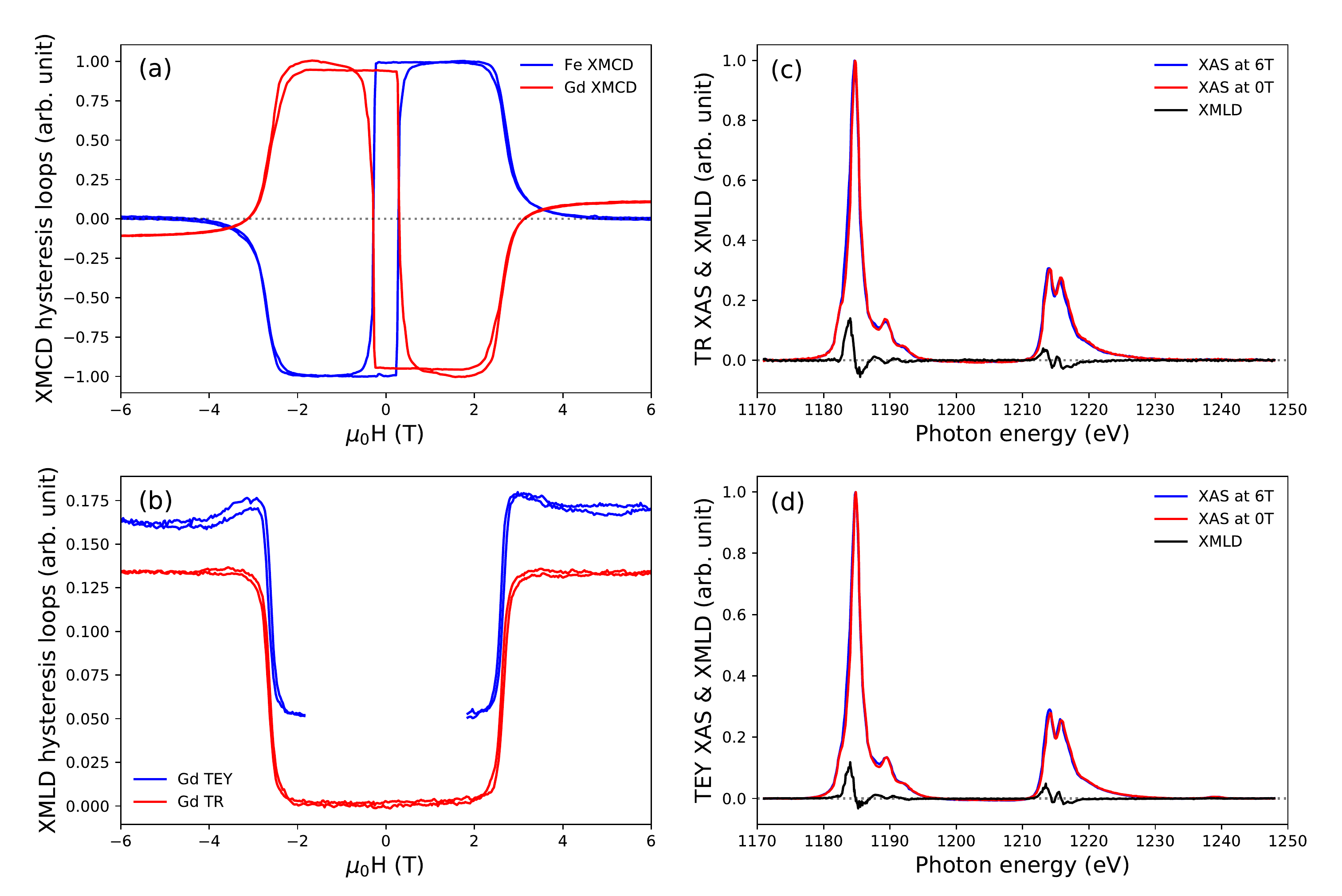}
	\caption{ (a) The out-of-plane XMCD hysteresis loops are measured by recording the transmission signals with \textbf{\textit{circular}} polarized X-rays at the Gd M$_{5}$ edge $E= 1184.9$ eV and at Fe L$_{3}$ edge $E= 707$ eV. They clearly demonstrate that the magnetic moments of the Gd and Fe are coupled anti-parallel one to each other. (b) The XMLD hysteresis loops are measured using the \textbf{\textit{linear}} polarized X-rays at the Gd M$_{5}$ edge $E= 1183.7$ eV. Note that, for the TEY loop the data at smaller fields are not shown due to a common artefact caused by the electrons near zero field during field sweep measurements, leading to large distortions of the curve. (c,d) The XAS and XMLD spectra measured at 0T and 6T. Here we observe a $\approx13.5\%$ XMLD signal for transmission (c) and $\approx10.6\%$ for TEY (d).}
	\label{fig:FeGd}
\end{figure}

In Figure~\ref{fig:FeGd} we provide an overview of the magnetic behaviour for a constant temperature equal to 35 K. The sample has been measured in transmission and TEY modes. The element specific (for Fe L$_3$ and Gd M$_5$ edges) magnetic hysteresis loops (Figure~\ref{fig:FeGd}(a)) were measured in transmission mode (bulk part) with circular polarized beam (XMCD) impinging perpendicular to the sample surface. They display a central rectangular loop which is characteristic for the perpendicular anisotropy, and the side loops. These hysteresis loops clearly show that the Fe and Gd sub-lattice magnetizations are anti-parallelly oriented with respect to one another. At the highest fields, one notices that a weak non-collinearity occurs. Strikingly, the regions between the -6~T and -2~ and between 2~T and 6~T are very similar to the so called transient ferromagnetic-like state observed in all optical switching experiments on FeCoGd samples~\cite{radu2011transient} (see also the Discussion section in the main manuscript).  

 The hysteresis loops with linear polarization (XMLD) for both the surface (TEY) and the bulk (transmission) parts exhibit a clear imprint of a large rotation of the sub-latices from out-of-plane direction to an in-plane direction (Figure~\ref{fig:FeGd}(b)). The differences between the surface and the bulk are also clearly observed, suggesting that the  surface spins  are more susceptible to the external fields. Finally, the XMLD spectra are shown for both TEY and transmission modes in Figure~\ref{fig:FeGd}(c, d). Similar to Dy, the Gd does also exhibit a large XMLD contrast which can be used for non-collinearity studies.
\newpage

\bibliographystyle{unsrt}
\bibliography{DyCo5}